\documentclass[10pt]{iopart}
\input epsf
\newcommand{\ba}{\begin{eqnarray}}
\newcommand{\ea}{\end{eqnarray}}

\begin{document}
\jl{4}
\title[Partial Dynamical SU(3) Symmetry]
{Partial Dynamical SU(3) Symmetry in Deformed Nuclei}
\author{Amiram Leviatan and Ilan Sinai}
\address{Racah Institute of Physics,
The Hebrew University, Jerusalem 91904,
Israel}

\begin{abstract}
We discuss characteristic features of $SU(3)$ 
partial dynamical symmetry in relation to nuclear spectroscopy and compare
with previous broken-$SU(3)$ calculations for $^{168}$Er.
\end{abstract}
\vspace{-0.8cm}

\vspace{12pt}
\noindent
Elliott's pioneering work \cite{ELL} established SU(3) as a relevant symmetry 
for axially-deformed nuclei.
In the Interacting Boson Model (IBM) \cite{IBM} of nuclei, such a symmetry 
appears as a dynamical symmetry in which rotational
bands are arranged into irreducible representations (irreps), 
$(\lambda,\mu)$, of SU(3). For $N$ bosons the ground band $g(K=0)$ spans 
the irrep $(2N,0)$ while the $\beta(K=0)$ and $\gamma(K=2)$ bands are 
degenerate and span the irrep $(2N-4,2)$. The undesired $\beta$-$\gamma$ 
degeneracy can be lifted by adding an $O(6)$ term to the $SU(3)$ Hamiltonian, 
as done by Warner, Casten and Davidson (WCD) \cite{WCD} or by considering
a general (non-SU(3)) quadrupole operator, as in the consistent-Q formalism
(CQF) \cite{CQF}. In these procedures the $SU(3)$ symmetry is completely 
broken, all eigenstates are mixed and none of the virtues of an
exact dynamical symmetry ({\it e.g.} solvability) 
are retained. In contrast, the recently introduced partial dynamical 
symmetry (PDS) \cite{LEV96} corresponds to a particular breaking of 
$SU(3)$, such that {\bf part} of the states (but not all) are
still solvable with good symmetry. 

The work of \cite{LEV96} demonstrated the relevance of 
$SU(3)$ PDS to the spectroscopy of $^{168}$Er. The Hamiltonian used has
the form
\ba
H \;=\; h_{0}P^{\dagger}_{0}P_{0} + h_{2}P^{\dagger}_{2}
\cdot\tilde P_{2} + \lambda\, \hat C_{O(3)} \quad ~. 
\ea
where 
$P^{\dagger}_{0} = d^{\dagger}\cdot d^{\dagger} - 2(s^{\dagger})^2$,
$P^{\dagger}_{2,\mu} =  2\,s^{\dagger}d^{\dagger}_{\mu} 
+ \sqrt{7}(d^{\dagger}d^{\dagger})^{(2)}_{\mu}$ are IBM boson-pairs
and $\hat C_{O(3)}$ the Casimir operator of $O(3)$.
For $h_{0}=h_{2}$ the first two terms of Eq. (1) form an SU(3) scalar 
related to the Casimir operator of $SU(3)$, while for $h_{0}=-5h_{2}$ they 
form an $SU(3)$ tensor, $(\lambda,\mu)=(2,2)$. Although
the Hamiltonian is not an $SU(3)$ scalar, it has a subset of 
{\bf solvable} states with good $SU(3)$ symmetry.
The $O(3)$ rotational term is diagonal, contributes an $L(L+1)$
splitting and has no effect on wave functions. 
The solvable eigenstates of $H$ belong to the ground and 
$\gamma^{k}_{K=2k}$ bands and are simply selected members of the 
Elliott basis \cite{ELL} with good $SU(3)$ symmetry, 
$(\lambda,\mu)=(2N-4k,2k)K=2k$.
States in other bands are mixed. 
The SU(3) decomposition of the lowest bands in $^{168}$Er is 
shown in Figure 1 and compared to the broken-SU(3) calculations of 
WCD \cite{WCD} and CQF \cite{CQF}. In the PDS calculation 
the states belonging to the ground and 
$\gamma$ bands are the Elliott states $\phi_{E}((2N,0)K=0,L)$ and 
$\phi_{E}((2N-4,2)K=2,L)$ respectively, while the $K=0_2$ band is mixed 
and has the structure
\ba
\fl
\vert L,K=0_2\rangle = A_1\tilde\phi_{E}((2N-4,2)\tilde K=0,L)
+ A_2\tilde\phi_{E}((2N-8,4)\tilde K=0,L)\nonumber\\
+ A_3\phi_{E}((2N-6,0)K=0,L) ~.
\ea
Here $\tilde\phi_{E}$ denote states orthogonal to the solvable $\gamma^k_{K=2k}$
Elliott's states. 
For $^{168}$Er ($N=16$), the $K=0_2$
band contains $9.6\%$ $(26,0)$ and 
$2.9\%$ $(24,4)$ admixtures into the dominant $(28,2)$ irrep.
Since the geometric analogs of the $SU(3)$ bands \cite{WC82} are 
$(2N-4,2)K=0 \sim \beta$, 
$(2N-8,4)K=0 \sim (\sqrt{2}\beta^2 + \gamma{^2}_{K=0})$, 
$(2N-6,0)K=0 \sim (\beta^2 - \sqrt{2}\gamma^{2}_{K=0})$, it follows
that in the PDS calculation, the $K=0_2$ band contains admixtures 
of $12.4\%$ $\gamma^{2}_{K=0}$ and $0.1\%$ $\beta^2$ into the $\beta$
mode. These observations are in line with recent theoretical \cite{CASBREN}
and experimental \cite{Lehmann} claims concerning the importance 
of a double-$\gamma$ component for the structure of the lowest $K=0$ 
excited band.

An important clue to the structure of these collective excitations
comes from E2 transitions. The relevant operator is
\ba
T(E2) \; = \; \alpha\, Q^{(2)} + \theta\, \Pi^{(2)}
\ea 
where $Q^{(2)}$ is the quadrupole $SU(3)$ generator and 
$\Pi^{(2)} = (\,d^{\dagger}s + s^{\dagger}\tilde d \,)$
is a (2,2) tensor under $SU(3)$. The parameters $\alpha$ and $\theta$
can be extracted from known B(E2) values for $2^{+}_g\rightarrow 0^{+}_g$
and $2^{+}_{\gamma}\rightarrow 0^{+}_g$ transitions. 
Since the wave functions of the solvable states are known, it is
possible to obtain {\bf analytic} expressions for the E2 rates
between them \cite{LEV96}. If we recall that only the ground
band has the $SU(3)$ component $(\lambda,\mu)=(2N,0)$, 
that $Q^{(2)}$, as a generator, cannot connect different $SU(3)$ irreps 
and that the $\Pi^{(2)}$ term can connect the $(2N,0)$ irrep only with 
the (2N-4,2) irrep, we obtain the
following expressions for B(E2) values of $K=0_2\rightarrow g$ and 
$\gamma\rightarrow g$ transitions
\begin{equation}
\eqalign{
\fl
B(E2;\gamma,L\rightarrow g,L') =
\theta^2\,{\vert\langle\phi_{E}((2N-4,2)K=2,L)||\Pi^{(2)}||
\phi_{E}((2N,0)K=0,L)\rangle\vert^{2}\over (2L+1)}\\
\fl
B(E2;K=0_2,L\rightarrow g,L') =\\
A_{1}^2\,\theta^2\,{\vert\langle\tilde\phi_{E}((2N-4,2)\tilde K=0,L)
||\Pi^{(2)}||\phi_{E}((2N,0)K=0,L)\rangle\vert^{2}\over (2L+1)}} \\
\end{equation}
The reduced matrix elements of $\Pi^{(2)}$ are known \cite{SU3,ISA}.
It follows that $B(E2)$ ratios for these transitions are independent of both 
$\alpha$ and $\theta$. Furthermore, since the
ground and $\gamma$ bands have pure $SU(3)$ character, 
the corresponding wave-functions do not depend
on parameters of the Hamiltonian and hence are determined solely by
symmetry. Consequently, the 
B(E2) ratios for $\gamma\rightarrow g$ transitions are 
parameter-free predictions of $SU(3)$ PDS. 
The latter were shown in \cite{LEV96} 
to be in excellent agreement with experiment and are similar 
to those of the WCD calculation \cite{WCD}.
The B(E2) values for $K=0_2\rightarrow g$ transitions
are seen from Eq. (4) to be proportional to $A_{1}^2$, hence, in accord
with the discussion following Eq. (2), determine the admixture 
of double-phonon excitations in the $K=0_2$ band. 
A comparison with recent data \cite{Lehmann} and previous broken-SU(3)
calculations \cite{WCD,CQF} is shown in Table 1.
The agreement between the PDS predictions and the data
confirms the relevance of partial dynamical $SU(3)$ symmetry to the 
spectroscopy of $^{168}$Er.
\noindent
This work is supported by a grant from the 
Israel Science Foundation.
\vfil\break
\section*{}

\begin{table}
\caption[]{Comparison of experimental and theoretical absolute B(E2) values 
[W.u.] for transitions from the $K=0_2$ band in $^{168}$Er.
\normalsize}
\vskip 10pt
 \begin{tabular}{lcc|ccc}
\hline
\multicolumn{3}{c|}{Exp. \cite{Lehmann}} &
\multicolumn{3}{c}{Calc.}\\
Transition & B(E2) & range & PDS & WCD \cite{WCD} & CQF \cite{CQF} \\
& & & & & \\
\hline
& & & & & \\
$2^{+}_{K=0_2}\rightarrow 0^{+}_g$ & 0.4 & 0.06--0.94 &
0.65  & 0.15  & 0.03  \\
& & & & & \\
$2^{+}_{K=0_2}\rightarrow 2^{+}_g$ & 0.5 & 0.07--1.27 &
1.02  & 0.24  & 0.03  \\
& & & & & \\
$2^{+}_{K=0_2}\rightarrow 4^{+}_g$ & 2.2 & 0.4--5.1 &
2.27 & 0.50 & 0.10 \\
& & & & & \\
$2^{+}_{K=0_2}\rightarrow 2^{+}_{\gamma}$ $^{a)}$ & 6.2 (3.1) 
& 1--15 (0.5--7.5) &
4.08 & 4.2 & 4.53 \\
& & & & & \\
$2^{+}_{K=0_2}\rightarrow 3^{+}_{\gamma}$ $^{a)}$ & 7.2 (3.6) 
& 1--19 (0.5--9.5) &
7.52 & 7.9 & 12.64 \\
& & & & & \\
\hline
\multicolumn{6}{l}{\small $^{a)}$ The two numbers in each entry
correspond to an assumption of pure E2}\\ 
\multicolumn{6}{l}{\small and (in parenthesis)
50\% E2 multipolarity.}\\
\end{tabular}
\end{table}

\newpage
\begin{figure}
\vspace{-2.5cm}
\hspace{-3.5cm}
\epsfbox{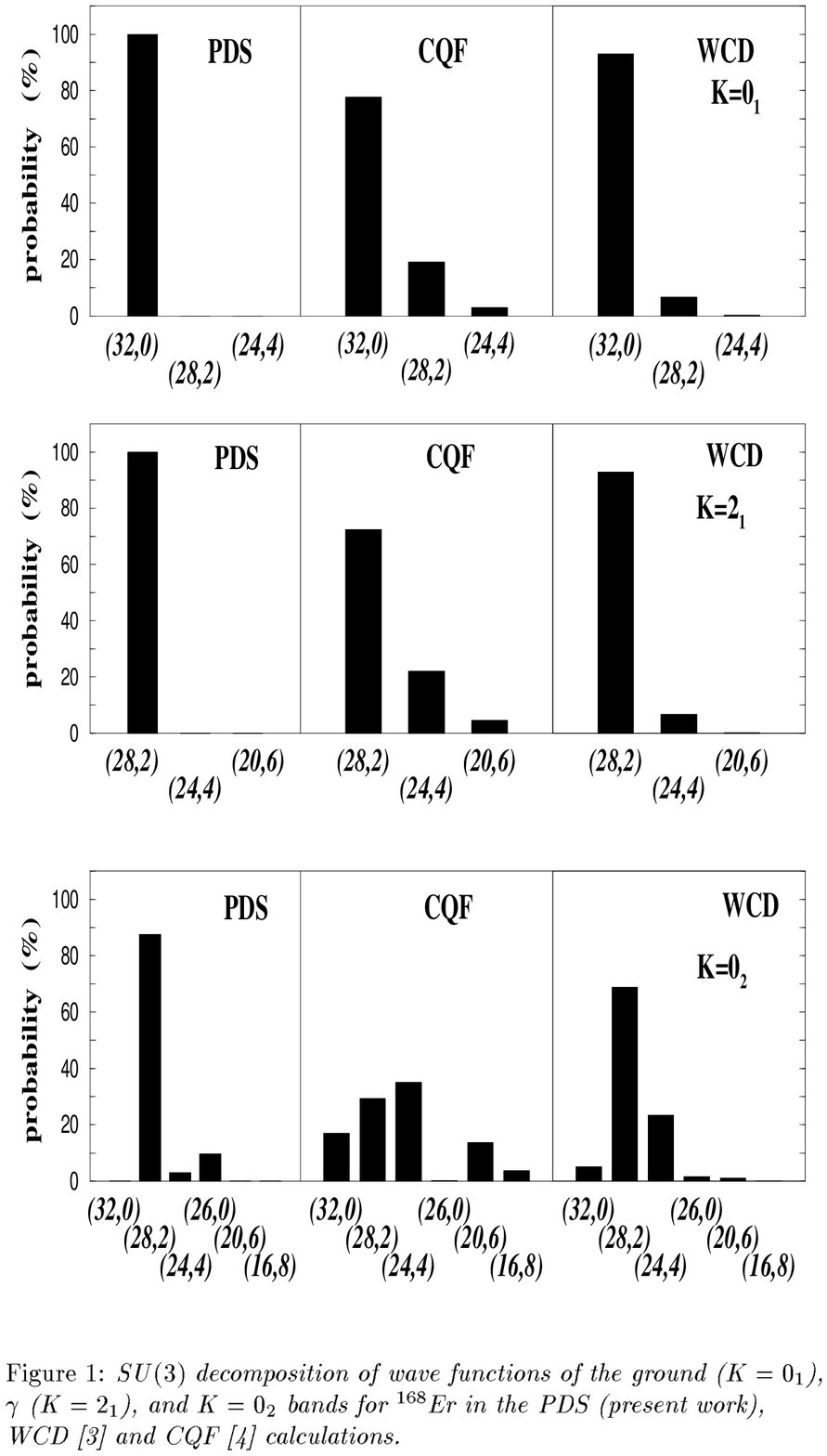}
\caption{
$SU(3)$ decomposition of wave functions of the ground ($K=0_1$)
\hfill\break
$\gamma$ ($K=2_1$), and $K=0_2$ bands of $^{168}$Er in the PDS 
\hfill\break
(present work), WCD \cite{WCD} and CQF \cite{CQF} calculations.}
\end{figure}

\end{document}